\documentclass[11pt]{article}
\usepackage{amsmath}
\usepackage{graphicx}
\newcommand{\sd}{\mathrm{d}}
\newcommand{\se}{\mathrm{e}}

\begin{document}
%\begin{flushright}$Revision: 1.33 $\hfill
%\verb$Date: 2006-09-25 05:39:25+09 $\end{flushright}

\begin{center}
{\LARGE\bf Species-area relationship for power-law species abundance distribution}\\
\vspace{10pt}{\large\bf Haruyuki Irie and Kei Tokita$^*$}\\
{\sl Information Media Center \& Grad. Scl. Sci., Hiroshima University\\
$^*$ Cybermedia Center, Grad. Scl. Sci. \& Grad. Scl. Frontier Biosci., Osaka University}
\end{center}

\vspace{12pt}
\noindent{\Large\bf Abstract}

\vspace{12pt}
We studied the mathematical relations between species abundance distributions (SADs) and species-area relationships (SARs) and found that a power-law SAR can be generally derived from a power-law SAD without a special assumption such as the ``canonical hypothesis''. In the present analysis, an SAR-exponent is obtained as a function of an SAD-exponent for a finite number of species. We also studied the inverse problem, from SARs to SADs, and found that a power-SAD can be derived from a power-SAR under the condition that the functional form of the corresponding SAD is invariant for changes in the number of species. We also discuss general relationships among lognormal SADs, the broken-stick model (exponential SADs), linear SARs and logarithmic SARs. These results suggest the existence of a common mechanism for SADs and SARs, which could prove a useful tool for theoretical and experimental studies on biodiversity and species coexistence.
\\


%\noindent\hrulefill
%%%%%%%%%%%%%%%%%%%%%%%%%%%%%%%%%%%%%%%%%%%%%%%%%%%%%%%%%%%%%%%%%
\newpage
\section{Introduction}

A fundamental question in ecology is how various species coexist in nature (Chave {\sl et al.} 2002; Hutchinson 1959; Levins 1970; May 1972; Pacala \& Tilman 1993; Rosenzweig 1995; Tokeshi 1999; Gaston \& Blackburn 2000; etc.). The answers to this question are expected to provide great insights into both theories of biodiversity and effective nature conservation practices.

Among the various explorations into the mechanisms of species coexistence, two community-level properties have been theoretically and quantitatively examined: species abundance distributions (SADs) (Motomura 1932; Fisher {\sl et al.} 1943; Preston 1948; MacArthur 1957; May 1975; Sugihara 1980; Harte {\sl et al.} 1999; Hubbell 2001) and species-area relationships (SARs) (Arrhenius 1921; Preston 1962a,1962b; MacArthur \& Wilson 1967; May 1975; Pueyo 2006). A mechanism to reproduce such macroscopic ecological patterns using a microscopic model has been one of the central issues in recent community ecology (Durrett \& Levin, 1996; Ney-Nifle \& Mangel, 1999; Tokita \& Yasutomi, 1999; Bastolla et al., 2001; Tokita \& Yasutomi, 2003; Tokita, 2004; Lawson \& Jensen, 2006; Tokita, 2006). Studies on these macroscopic patterns, therefore, not only give theoretical insight into large-scale ecosystems but also clarify the impacts of habitat fragmentation. Thus, they aid in efforts to devise long-term estimations and strategies for nature conservation.

SAD and SAR are mutually connected to each other. Preston (1962) derived the power-law SAR from the lognormal SAD under an assumption called the canonical hypothesis, which states that the peak of the individuals curve coincides with the number of individuals in the most abundant species. May (1975) comprehensively studied various types of SAD such as Preston's lognormal distribution, uniform distribution, MacArthur's Broken Stick (exponential) distribution, Motomura's geometric series distribution and Fisher's logseries distribution, and demonstrated that the first three SADs lead to power-law SAR and the latter two correspond to log SAR.

In addition to those pioneering works on SADs, power-law SAD has been reported (Margalef 1994; Pueyo 2006). Other than SADs, power-law has been observed for relationships between abundance and body size (Siemann {\sl et al.}  1996).  Power-law is, therefore, ubiquitous in biology. It is moreover known that, in general, power-law distribution is given in a limit of large variance of the lognormal distribution.

Here, we demonstrate that power-law SAR can be mathematically derived from power-law SAD without any such assumption as the canonical hypothesis. We also discuss an inverse problem: namely, what type of SAD can be obtained when we start from the power-law SAR?

%%%%%%%%%%%%%%%%%%%%%%%%%%%%%%%%%%%%%%%%%%%%%%%%%%%%%%%%%%%%%%%%%
%\newpage
\section{Species abundance distribution and rank}

Let $\sigma(x)\sd x$ be the species abundance distribution (SAD) between the number of individuals $x\sim x+\sd x$. The total number of species $S$ is obtained by integrating $\sigma(x)$ from the minimum value of $x=m$ to the maximum value of $x=X$ as
\begin{equation}\label{eq.S}
	S=\int_m^X\sigma(x)\sd x.
\end{equation}
The species rank of the number of individuals $x$ is defined by
\begin{equation}\label{eq.R}
	R(x)=\int_x^\infty\sigma(x')\sd x'.
\end{equation}
The inverse function of eqn (\ref{eq.R}), $x_R$, is the rank-abundance distribution; that is, the number of individuals of the $R$-th rank is $x_R$. The first rank of species, i.e. the most abundant species, has the maximum number of individuals, $X$; then we obtain 
\begin{equation}	\label{eq.RX}
	R(X)=1.
\end{equation}
This equation (\ref{eq.RX}) is equivalent to the estimation used by Preston (1962a,1962b), and May (1975).

Using the SAD $\sigma(x)$, the total population or the total number of individuals becomes
\begin{equation}\label{eq.N}
	N=\int_m^X x\sigma(x)\sd x.
\end{equation}
Writing the number of individuals normalized by the minimum value $m$ as $\hat{x}=x/m, \hat{N}=N/m$, and $\hat{\sigma}(\hat{x})\sd \hat{x}=\sigma(x)\sd x$, we obtain
\begin{equation*}
	S=\int_1^{\hat{X}} \hat{\sigma}(\hat{x})\sd \hat{x},\quad
	R=\int_{\hat{x}}^\infty\hat{\sigma}(\hat{x}')\sd\hat{x}',\quad
	\hat{N}=\int_1^{\hat{X}}\hat{x}\hat{\sigma}(\hat{x})\sd\hat{x}.
\end{equation*}

Hereafter, we omit the hat $\hat{}$, and we consider $x, X, \sigma$, and $N$ to be the normalized quantities.

%%%%%%%%%%%%%%%%%%%%%%%%%%%%%%%%%%%%%%%%%%%%%%%%%%%%%%%%%%%%%%%
%\newpage
\section{Power-law SAD}

%---------------------------
\subsection{Power-law SAR}

We consider here an SAD which decays with power-law from the minimum number of individuals, $x=1$ (the number normalized by the minimum value), to the maximum value, $x=X$ as
\begin{equation}	\label{eq.powsad}
	\sigma(x)=\tilde{S}\alpha x^{-(1+\alpha)},
\end{equation}
where $\tilde{S}$ is a constant. The rank function $R(x)$ is calculated by integrating $\sigma (x)$ from $x$ to infinity
\begin{equation}\label{eq.rank}
	R(x)=\int_x^\infty\sigma(x')\sd x'
	=\tilde{S}\alpha\left.\frac{x'^{-\alpha}}{-\alpha}\right|_x^\infty
	=\tilde{S}x^{-\alpha}.
\end{equation}
Using this, we find
\begin{equation}
\tilde{S}\equiv \int_{1}^{\infty}\sigma(x)\sd x = \int_{1}^{X}\sigma(x)\sd x + \int_{X}^{\infty}\sigma(x)\sd x=S+1.
\end{equation}
where eqns (\ref{eq.S}) and (\ref{eq.RX}) are used for the last equality.

The relation of the total number of species $S$ and the maximum number of individuals $X$ is obtained by eqns (\ref{eq.RX}) and (\ref{eq.rank}), $1=R(X)=\tilde{S}X^{-\alpha}$, so we obtain
\begin{equation}\label{eq.X}
	X^\alpha=\tilde{S}=S+1.
\end{equation}

By inserting eqn (\ref{eq.powsad}) into eqn (\ref{eq.N}), the total number of individuals becomes
\begin{equation}	\label{eq.NX}
	N=\int_1^X x\sigma(x)\sd x=\alpha\frac{X-X^\alpha}{1-\alpha}.
\end{equation}
For very large $X$, from eqn (\ref{eq.NX}), $N\simeq\alpha X/(1-\alpha)$ and $\tilde{S}\propto N^{\alpha}$ for $\alpha<1$, and $N\simeq\alpha X^\alpha/(\alpha-1)$ and $\tilde{S}\propto N$ for $\alpha>1$. When the size of area $A$ is proportional to the normalized total population $N$, the SAR becomes $S=cA^z$ with $z=\alpha$ for $\alpha < 1$ and $z=1$ for $ \alpha > 1$.

For a finite value of $X$, the SAR never becomes a simple power-law relation. In this case, let us start to define the SAR-exponent $\zeta$ as
\begin{equation}
	\zeta\equiv\frac{\sd\ln S}{\sd\ln A}.
\end{equation}
This can quantify the increasing rate of the species richness with increasing area size $A$. For the power SAR, $S=cA^z$, this exponent becomes $\zeta=z$, and for the logarithmic case, $S=K\ln A+a$, $\zeta=1/(\ln A+a/K)$. This exponent is closely related to the persistence function $a(A)$ introduced by Plotkin {\sl et al.} (2000) as $\zeta=-\log a(A)$.

If the normalized total population $N$ is proportional to the area size $A$, this SAR exponent equals $\zeta=\sd\ln S/\sd\ln N$. In the case of the power-law SAD eqn (\ref{eq.powsad}), we obtain
\begin{equation}	\label{eq.zeta}
	\zeta=\frac{\sd y}{\sd\mu}
	=\frac{1-\se^{-\theta y}}{1+\theta-\se^{-\theta y}}
	=\frac{1}{1+\displaystyle\frac{\theta}{1-\se^{-\theta y}}},
\end{equation}
where $\mu\equiv\ln N,\ y\equiv\ln \tilde{S}\simeq\ln S$ using $S\simeq \tilde{S}\gg 1$ for a large community, and $\theta\equiv\frac{1}{\alpha}-1$. For large $N$ and $X$ and in the limit of an infinite number of species ($S\to\infty$), $\zeta$ becomes constant:
\begin{equation}
	\zeta=\begin{cases}\alpha & (\mbox{for }\alpha<1)\\
		1 & (\mbox{for }\alpha\ge 1)\end{cases}.
\end{equation}
In Fig. 1, we show the SAR exponent $\zeta$ v.s. $\alpha$. The $\zeta$ bends near $\alpha=1$.

%%%%%%%%%%%%%%% Fig.1 %%%%%%%%%%%%%%%
\begin{figure}[!ht]\begin{center}
\includegraphics[width=.6\linewidth]{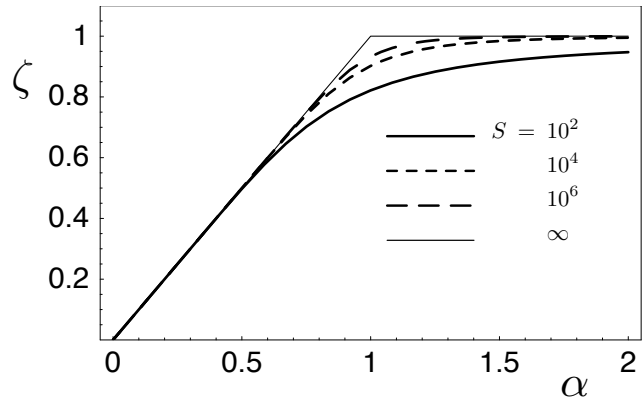}
\caption{SAR exponent $\zeta$ v.s. the exponent $\alpha$ for the power-law SAD eqn (\ref{eq.powsad})}
\end{center}\end{figure}

%%\vspace{1cm}
%%\noindent\hrule
%%\begin{center} Fig. 1\end{center}
%%\noindent\hrule
%%\vspace{1cm}
%%%%%%%%%%%%%%%%%%%%%%%%%%%%%%%%%%%%

%---------------------------
\subsection{Logarithmic SAR}

For $\alpha\to 0$, keeping $S_{0}\equiv \tilde{S}\alpha$ constant, the SAR becomes logarithmic, because for the SAD, $\sigma(x)=S_0 x^{-(1+\alpha)}$ and 
\begin{eqnarray}
  S &=& \int_{1}^{X}\sigma(x)\sd x
	= \left.-\frac{S_{0}x^{-\alpha}}{\alpha}\right|_{1}^{X}
	=S_{0}\frac{1-X^{-\alpha}}{\alpha}
	\to S_{0}\left.\frac{\partial_{\alpha}(1-\exp(-\alpha\ln X))}
	{\partial_{\alpha}\alpha}\right|_{\alpha=0}\nonumber\label{eq.S-S0lnx}\\
	&=&S_{0}\ln X,\\
  N &=& \int_{1}^{X}x\sigma(x)\sd x
	=S_{0}\int_{1}^{X}x^{-\alpha}\sd x
	=\left.\frac{S_{0}}{1-\alpha}x^{1-\alpha}\right|_{1}^{X}
	=\frac{S_{0}}{1-\alpha}(X^{1-\alpha}-1)\nonumber\\
	&\to& S_{0}X\quad (\mbox{for}\, \alpha\to 0),
\end{eqnarray}
where $\partial_{\alpha}$ denotes the derivative in $\alpha$, and L'h\^{o}pital's rule is used in taking the limit in eqn \ref{eq.S-S0lnx}; then we derive
\begin{equation}
	\frac{S}{S_0} \sim \ln \frac{N}{S_0}.
\end{equation}
This case $\alpha\to 0$ corresponds to the continuous version of the geometric SAD (Motomura 1932, May 1975). In the geometric SAD, the rank-size distribution is $x_i=NC_k k(1-k)^{i-1}$ with a constant $k$, and the coefficient $C_k$ being the normalization constant given by the condition $\sum_{i=1}^S x_i=N$, $C_k=1/[1-(1-k)^S]$. The inverse function of this expression of $x_i$ leads to the rank function by setting $i=R$,
\begin{equation}
	i=R(x)=1+K\ln(NC_k k)-K\ln x,\quad
	\left(K\equiv\frac{1}{\ln\frac{1}{1-k}}\right).
\end{equation}
Differentiating $R(x)$ by $x$, we obtain
\begin{equation}	\label{eq.geosad}
	\sigma(x)=-\frac{\sd R(x)}{\sd x}=\frac{K}{x}.
\end{equation}
If we express it by the normalized quantities, $X=x_1=NC_k k, 1=x_S=NC_k k(1-k)^S$. For large $N, S$, and $X$, the parameter $C_k$ becomes $C_k\to 1$, and $X\sim kN, S=K\ln (kN)$; finally, we obtain the logarithmic SAR. From eqn (\ref{eq.geosad}), we find that the SAD corresponds to the power-law SAD eqn (\ref{eq.powsad}) with $\alpha\to 0$, and we can obtain the logarithmic SAR for the power-law SAD with $\alpha\to 0$ in the continuous approximation as well. The relation between parameters $S_0,\ K$ and $k$ is $S_0=K\sim 1/k$.

%%%%%%%%%%%%%%%%%%%%%%%%%%%%%%%%%%%%%%%%%%%%%%%%%%%%%%%%%%%%%%%
%\newpage
\section{From SAR to SAD}

In contrast to the previous section, we consider here the inverse problem of what kind of SAD can be derived from a given SAR. If the shape of the SAD is unchanged with increasing $S, X$, and $N$, we can write $\sigma(x)=\tilde{S}p(x)$: only the coefficient $\tilde{S}$ varies with varying $S$, but the function $p(x)$ is unchanged. In this case, eqn (\ref{eq.S}) becomes
\begin{equation}	\label{eq.S2}
	S=\tilde{S}\int_1^X p(x)\sd x,
\end{equation}
with the use of the normalized quantities (normalized by the minimum number of individuals). Defining the cumulative distribution function as 
\begin{equation}	\label{eq.P}
	P(x)=\int_x^\infty p(x')\sd x' \quad \left(=\frac{R(x)}{\tilde{S}}\right),
\end{equation}
we obtain
\begin{eqnarray}\label{eq.PX}
	\frac{1}{\tilde{S}}=P(X)=\int_X^\infty p(x)\sd x,
\end{eqnarray}
from eqns (\ref{eq.R}) and (\ref{eq.RX}).

From eqn (\ref{eq.N}), the normalized total population $N$ divided by $\tilde{S}$ is
\begin{equation}	\label{eq.NS}
	\frac{N}{\tilde{S}}=\int_1^X xp(x)\sd x.
\end{equation}
If $\int_1^\infty p(x)\sd x=1$, from eqns (\ref{eq.S2}) and (\ref{eq.PX}), we obtain
\begin{equation}
	\tilde{S}=\tilde{S}\int_1^\infty p(x)\sd x=\tilde{S}\int_1^Xp(x)\sd x+\tilde{S}\int_X^\infty p(x)\sd x
	=\tilde{S}+1\simeq \tilde{S}.
\end{equation}
Equations (\ref{eq.PX}) is the relation between $X$ and $S$ through $P(X)$ on the one hand, and eqn (\ref{eq.NS}) is the relation between $N$ and $S$ on the other.

The SAR is the relation between $N$ and $S$ if $N\propto A$, and if we put $N=S F(1/S)$, from eqn (\ref{eq.PX}), $N/S=F(P(X))$. Therefore, if $X$ is changed to $X+\delta X$, the variation of eqn (\ref{eq.NS}) becomes
\begin{equation}	\label{eq.var}
	\delta F(P(X))=-X\delta P(X)
		+\int_1^X x\frac{\partial p(x;\beta)}{\partial\beta}\sd x\delta\beta,
\end{equation}
where $\beta$ expresses some parameters of the distribution, and we use $\delta P(X)=-p(X)\delta X$. Because we consider the case in which the parameters are unchanged with varying $N, S$, and $X$, the last term vanishes. For example, if the SAR is power-law $S=cN^z$, $N=c'S^{1/z}=c'P(X)^{-1/z}$; that is, $F(P)=c'P^{1-1/z}$. Substituting this expression into eqn (\ref{eq.var}), we obtain
\begin{equation}
	c'\left(1-\frac{1}{z}\right)P(X)^{-1/z}\delta P(X)=-X\delta P(X).
\end{equation}
Therefore, the exponent $z$ must satisfy
\begin{equation}
	z < 1 \quad\mbox{and}\quad P(X)\propto X^{-z}.
\end{equation}
This means that the tail of the SAD is a power-law of eqn (\ref{eq.powsad}), and the exponent becomes $\alpha=z<1$.

%%%%%% Discussion %%%%%%%%%%%%%%%%%%%%%%%%%%%%%%%%%%%%%%%%%%%%%%%%%%%%%%%%%
%\newpage
\section{Discussion}

We obtained the power-law SAR for a power-law SAD using the classical method of R. May and Preston. We also considered the inverse problem of obtaining the power-law SAD from a given power-law SAR if the shape of the SAD is unchanged with varying total population, area size, and species richness.

R. May (1975) obtained the power-law SAR for the lognormal SAD and the broken stick model SAD. In both cases, the parameters of SAD vary with increasing $N$: in the lognormal SAD, the variance of the log of the number of individuals increases proportional to $\ln X$, and for the case of the broken stick model, in which the SAD becomes an exponential distribution, the SAR is obtained if the average of the number of individuals is proportional to the total number of species $S$. The last term of eqn (\ref{eq.var}) does not vanish in either SAD.

In the case of a linear SAR, $S=cN$, first we consider that this linearity holds in the limit of $S\to\infty$ and assume that any deviation from the linearity becomes $N/S=1/c-bS^{-\gamma}\, (\gamma>0)$, then from eqn (\ref{eq.var}) $b\gamma P^{\gamma-1}\delta P=X\delta P$ and $P(X)\propto X^{-1/(1-\gamma)}$. Hence, this case corresponds to the power-law SAD with $\alpha=1/(1-\gamma)>1$. If the linearity holds completely for all $S$ as $b=0$ above, the variation of the maximum number of individuals becomes zero, $\delta X=0$, and this corresponds to the case in which $X$ is constant, for example, a uniform SAD (May 1975), $\sigma(x)\propto \delta(x)$.

We found in the previous section that the logarithmic SAR, $S=K\ln N + a$, is derived from the case of the SAD$\propto 1/x$, and the rank function defined by eqn (\ref{eq.R}) diverges, so we cannot use eqn (\ref{eq.var}) as a method for deriving SAD from SAR.

Harte {\sl et al.} (1999) obtained an SAD from a power-law SAR using the renormalization group technique for the case of existing self-similarity: the fraction of a species found in an area with a size $A$, which is also found in $A/2$, is independent of $A$ and the abundance $x$. But this SAD is far from the power-law SAD. Pueyo (2006) pointed out that there are possibilities for other shapes of SADs if the fraction depends on the abundance, and he also obtained power-law SAD from more straightforward discussion.

\vspace{10mm}
\noindent{\bf\Large Acknowledgements}

\vspace{12pt}
The authors thank T. Chawanya for fruitful discussions. The authors also thank the members of Department of Mathematical and Life Sciences, Graduate School of Science, Hiroshima University and the members of Large-scale Computational Science Division (Kikuchi lab), Cybermedia Center, Osaka University. The present study has been supported by Grants-in-Aid from MEXT, Japan (17540383 and Priority Areas "Systems Genomics") and by the research fund of Fukken Co., Ltd..

\vspace{24pt}
%%%%%%% References %%%%%%%%%%%%%%%%%%%%%%%%%%%%%%%%%%%%%%%%%%%%%%%%%%%%%%%%
%\newpage
%----------------
\newlength{\IN} \IN=20pt
\newenvironment{Rfrnc}{\par
	\addtolength{\leftskip}{\IN}
	\setlength{\parindent}{-\IN}
\begingroup}{\endgroup\par}
%----------------

%\vspace{10mm}
\noindent{\bf\Large References}

\vspace{12pt}
\begin{Rfrnc}
%\hspace*{-\IN}%\hspace{-3pt}
%------------------
Arrhenius, O. (1921).
Species and area.
{\em J. Ecol.}, 9, 95--99.

Bastolla, U., Lassig, M., Manrubia, S., and Valleriani, A. (2001).
Diversity patterns from ecological models at dynamical equilibrium.
{\em J. Theor. Biol.}, 212, 11--34.

Chave, J., Muller-Landau, H., \& Levin, S.A. (2002).
Comparing classical community models, theoretical consequences for patterns of diversity.
{\em Am. Nat.}, 159, 1--23.

Durrett, R. and Levin, S. (1996).
Spatial models for species-area curves.
{\em J. Theor. Biol.}, 179, 119--127.

Fisher, R., Corbet, A., \& Williams, C. (1943).
The relation between the number of species and the number of individuals in a random sample of an animal population.
{\em J. Anim. Ecol.}, 12, 42--58.

Gaston, K. \& Blackburn, T. (2000).
{\em Patterns and processes in macroecology}.
Blackwell Science.

Harte, J., Kinzig, A., \& Green, J. (1999).
Self-similarity in the distribution and abundance of species.
{\em Science}, 284, 334--336.

Hubbell, S. (2001).
{\em The Unified Neutral Theory of Biodiversity and Biogeography}.
Princeton Univ. Press.

Hutchinson, G. (1959).
Homage to santa rosalia, or why are there so many kinds of animals?
{\em Am. Nat.}, 93, 145--159.

Lawson, D. and Jensen, H. (2006).
The species-area relationship and evolution.
{\em J. Theor. Biol.}

Levins, R. (1970).
Extinction.
in {\em Some mathematical questions in biology},
Gerstenhaber, M., ed., vol.3, 75--108.
American Mathematical Society.

MacArthur, R. (1957).
On the relative abundance of bird species.
{\em Proc. Natl. Acad. Sci. U.S.A.}, 43, 293--295.

MacArthur, R. \& Wilson, E. (1967).
{\em The theory of island biogeography}.
Princeton Univ. Press.

Margalef, R. (1994).
Through the looking glass, how marine phytoplankton appears through the microscope when graded by size and taxonomically sorted
{\em Sci. Mar.}, 58, 87--101.

May, R.M. (1972).
Will a large complex system be stable?
{\em Nature}, 238, 413--414.

May, R.M. (1975).
Patterns of species abundance and diversity. In: {\em Ecology and evolution of communities} (eds. Cody, M.L. \& Diamond, J.M.).
Belknap Press of Harvard University Press, pp. 81--120.

Motomura, I. (1932).
On the statistical treatment of communities.
{\em Zool. Mag., Tokyo (in Japanese)}, 44, 379--383.

Ney-Nifle, M. and Mangel, M. (1999).
Species-area curves based on geographic range and occupancy.
{\em J. Theor. Biol.}, 196, 327--342.

Pacala, S. \& Tilman, D. (1993).
Limiting similarity in mechanistic and spatial models of plant competition in heterogeneous environments.
{\em Am. Nat.}, 143, 222--257.

Plotkin, J., Potts, M., Hubbell, S., \& Nowak, M. (2000).
Predicting species diversity in tropical forests.
{\em Proc. Natl. Acad. Sci. U.S.A.}, 97, 10850.

Preston, F. (1948).
The commonness and rarity of species.
{\em Ecology}, 29, 254--283.

Preston, F. (1962a).
The canonical distribution of commonness and parity,  Part i.
{\em Ecology}, 43, 185--215.

Preston, F. (1962b).
The canonical distribution of commonness and parity,  Part ii.
{\em Ecology}, 43, 410--432.

Pueyo, S. (2006).
Self-similarity in species-area relationship and in species abundance distribution.
{\em Oikos}, 112, 156--162.

Rosenzweig, M. (1995).
{\em Species Diversity in Space and Time}.
Cambridge Univ. Press.

Siemann, E., Tilman, D., \& Haarstad, J. (1996).
Insect species diversity, abundance and body size relationships.
{\em Nature}, 380, 704--706.

Sugihara, G. (1980).
Minimal community structure, an explanation of species abundance patterns.
{\em Am. Nat.}, 116, 770--787.

Tokeshi, M. (1999).
{\em Species coexistence -- ecological and evolutionary perspectives}.
Blackwell Science.

Tokita, K. (2004).
Species abundance patterns in complex evolutionary dynamics.
{\em Phys. Rev. Lett.}, 93, 178102.

Tokita, K. (2006).
Statistical mechanics of relative species abundance.
{\em Ecol. Informatics}, 1 (Available online 7 July, 2006).

Tokita, K. and Yasutomi, A. (1999).
Mass extinction in a dynamical system of evolution with variable dimension.
{\em Phys. Rev. E}, 60, 842--847.

Tokita, K. and Yasutomi, A. (2003).
Emergence of a complex and stable network in a model ecosystem with extinction and mutation.
{\em Theor. Popul. Biol.}, 63, 131--146.

\end{Rfrnc}

%%%%%% Figure Captions %%%%%%%%%%%%%%%%%%%%%%%%%%%%%%%%%%%%%%%%%%%%%%%%%%%%%%
\newpage
%%\noindent{\bf\large Figure Captions}

%%\noindent{Fig. 1. }
%%SAR exponent $\zeta$ v.s. the exponent $\alpha$ for the power-law SAD eqn (\ref{eq.powsad})

\end{document}